\documentclass[sigconf,authorversion=true]{acmart}
\AtBeginDocument{%
  \providecommand\BibTeX{{%
    \normalfont B\kern-0.5em{\scshape i\kern-0.25em b}\kern-0.8em\TeX}}}

\setcopyright{acmcopyright}
\copyrightyear{2023}
\acmYear{2023}
\acmDOI{XXXXXXX.XXXXXXX}
\acmDOI{}

\acmConference[ISPD 2023]{International Symposium on Physical Design}{March 26--29, 2023}{Online}
\newcommand{\smallerspace}{\vspace{-0.8em}}
\newcommand{\littlesmallerspace}{\vspace{-0.5em}}    
\usepackage{soul}
\usepackage{multirow}
\usepackage{ragged2e}
\usepackage[noend]{algpseudocode}     
\usepackage{algorithm}            
\usepackage{threeparttable}
\usepackage{subfig}
\algrenewcommand\algorithmicforall{\textbf{foreach}}

\begin{document}

%%%
%%% The "author" command and its associated commands are used to define
%%% the authors and their affiliations.
%%% Of note is the shared affiliation of the first two authors, and the
%%% "authornote" and "authornotemark" commands
%%% used to denote shared contribution to the research.
%%%
\author{Saideep Sreekumar, Mohammed Ashraf, Mohammed Nabeel, Ozgur Sinanoglu, Johann Knechtel}
\email{{sds710, ma199, mtn2, ozgursin, johann}@nyu.edu}
\affiliation{%
  \country{New York University Abu Dhabi, UAE}
  %\streetaddress{P.O. Box 1212}
  %\country{UAE}
}


%%% A "teaser" image appears between the author and affiliation
%%% information and the body of the document, and typically spans the
%%% page.
%  \includegraphics[width=\textwidth]{template/sampleteaser}

\newcommand{\jk}[1]{{\color{orange}JK: {#1}}}

\newcommand{\T}[2]{\#TTD(${#1}\%$, {#2})}

\title{X-Volt: Tuning Driver Strengths and Supply Voltages to Defend Power Side-Channel Attacks}
\title{X-Volt:
		Tuning Driver Strengths and Supply Voltages for an\\
		Efficient Defense against Power Side-Channel Attacks}
\title{X-Volt:
		Joint Tuning of Driver Strengths and Supply Voltages Against Power Side-Channel Attacks}

\begin{abstract}
Power side-channel (PSC) attacks are well-known threats to sensitive hardware like
advanced encryption standard (AES) crypto cores.
Given the significant impact of supply voltages (VCCs) on power profiles, various countermeasures 
based on VCC tuning have been proposed, among other defense strategies.
Driver strengths of cells, however, have been largely overlooked, despite having direct and significant impact
on power profiles as well.

For the first time, we thoroughly explore
the prospects of jointly tuning driver strengths and VCCs as novel working principle for PSC-attack countermeasures.
Toward this end, we take the following steps:
1)~we develop a simple circuit-level scheme for tuning;
2)~we implement a CAD flow for design-time evaluation of
ASICs,
enabling security assessment of ICs before tape-out;
3)~we implement a correlation power analysis (CPA) framework for thorough and comparative security analysis;
4)~we conduct an extensive experimental study of a regular AES design, implemented in ASIC as well as FPGA fabrics, under
various tuning scenarios;
5)~we summarize design guidelines for secure and efficient joint tuning.

In our experiments, we observe that runtime tuning is more effective than static tuning, for both ASIC and FPGA
implementations. For the latter, the AES core is rendered >11.8x (i.e., at least 11.8 times) as resilient as the
untuned baseline design.
Layout overheads can be considered acceptable, with, e.g., around +10\% 
critical-path delay for the most resilient tuning scenario in FPGA.

We will release source codes for our methodology, as well as artifacts from the experimental study, post peer-review.

\end{abstract}

\maketitle

\section{Introduction}
\label{sec:intro}

\textbf{Background:}
To protect sensitive data handled within integrated circuits (ICs), the use of cryptographic (crypto) modules is widely
adopted. Such modules are
based on provably secure algorithms for encryption/decryption of data.  Still, once attackers have access to ICs, direct or
even only remote/indirect, they can monitor
the runtime behaviour and physical interactions with the environment, e.g.,
via measurements (direct) or via software interfaces to embedded sensors (remote/indirect).
Such observations enable so-called \textit{side-channel attacks}~\cite{zhou05}, which can
serve to infer the secret key used for crypto modules, etc.

\textbf{Limitation of Prior Art:}
Power side-channel (PSC) attacks are a well-known and effective type of side-channel
attacks~\cite{brier04,skorobogatov12}.
Thus, a plethora of PSC countermeasures have been proposed, e.g., masking and hiding~\cite{li17_SCA%,9015513
}, voltage
switching~\cite{8361769,9006696%,10.1145/2345770.2345774%,DBLP:journals/corr/abs-1907-09440
}, noise injection~\cite{bellizia18%,8351968
}, etc.
Given the direct impact of supply voltages (VCCs) on power profiles, various countermeasures are
based on some kind of VCC tuning.
Driver strengths of cells, however, have been largely overlooked, despite significant
impact on power profiles as well.

\begin{figure}[tb]
\includegraphics[width=.85\columnwidth]{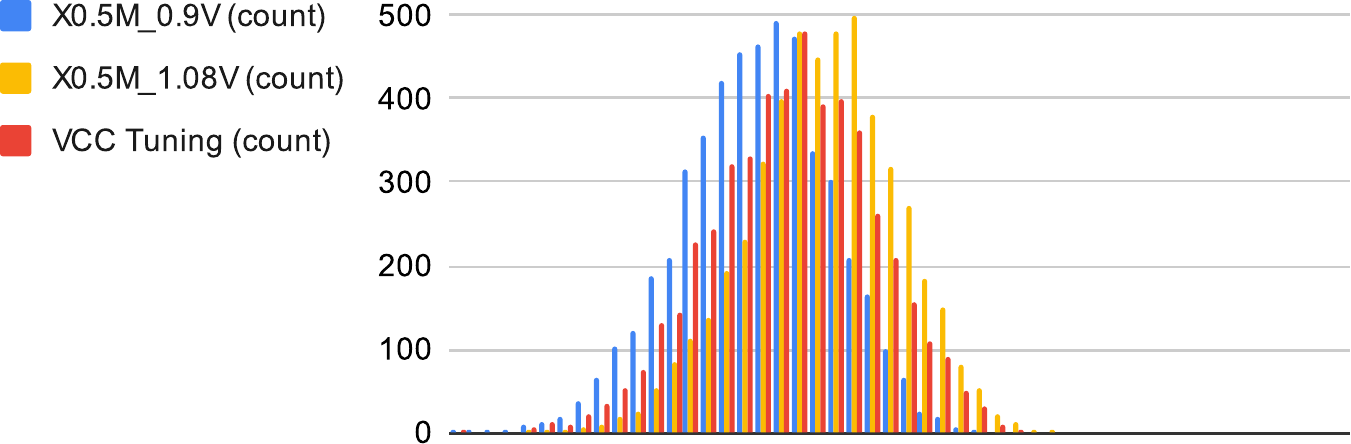}\\[3mm]
\includegraphics[width=.85\columnwidth]{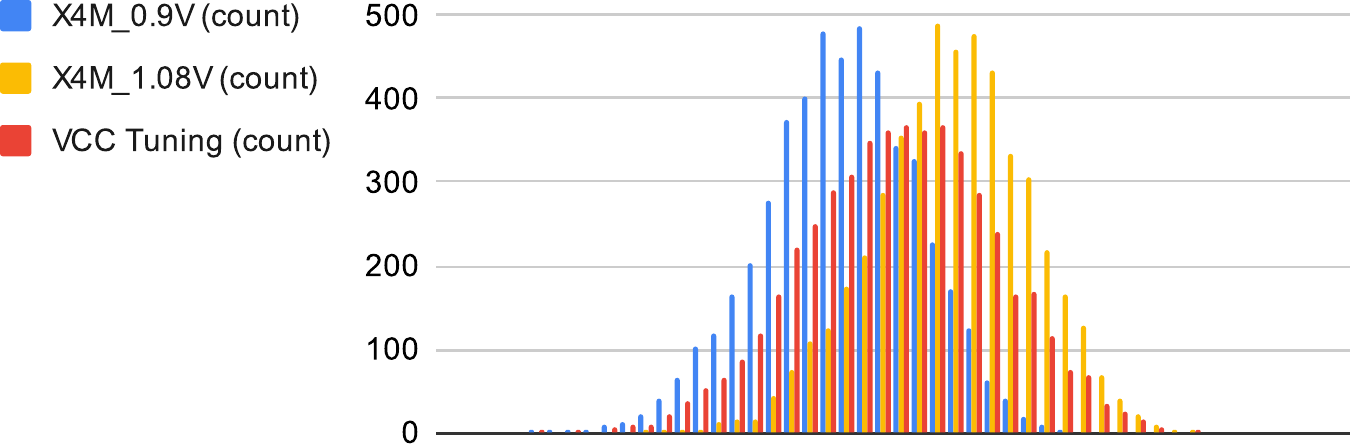}\\[3mm]
\includegraphics[width=\columnwidth]{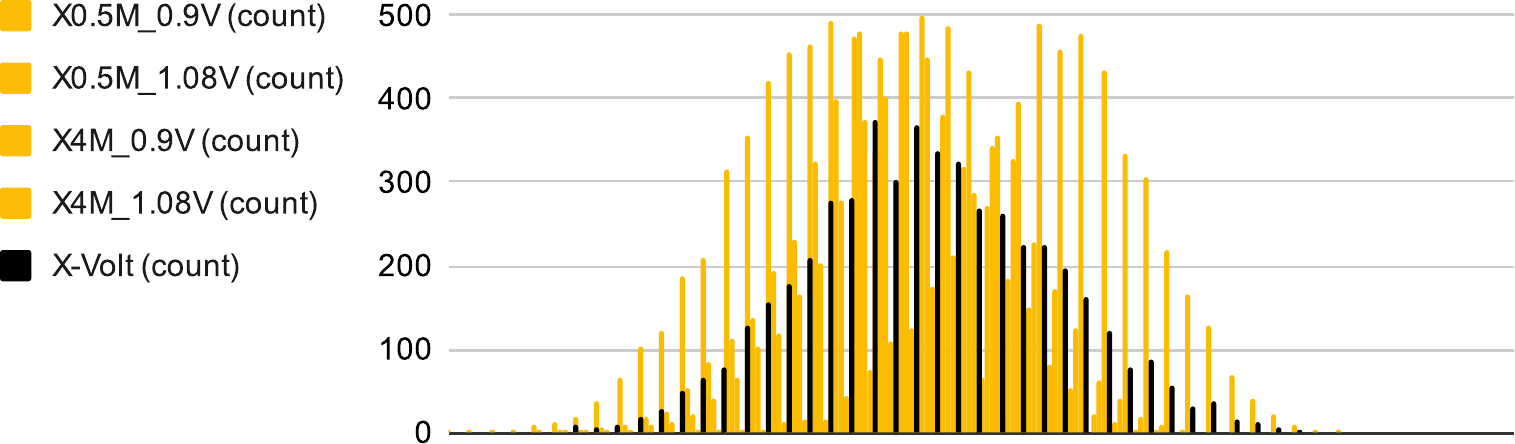}\\[3mm]
\smallerspace
	\caption{Motivational example for the impact of tuning.
		Histograms of power profiles, for a regular AES core implemented in a GlobalFoundries 55nm
		technology.
			The two ``VCC Tuning'' scenarios at the top are for
			two different cases of driver strengths assigned to all AES
			  flip-flops. These scenarios demonstrate two aspects of tuning:
			  (i) VCC tuning results in power profiles (red) that largely overlap with
			  the baseline profiles (blue and yellow) and (ii) the overlap or rather shape/distribution of
			  the tuned profile depends on the driver strengths.
			The ``X-Volt'' scenario at the bottom demonstrates how joint tuning of driver strengths and VCCs renders the
			resulting profile (black) even more interspersed.}
	\label{fig:power_profiles}
\smallerspace
\end{figure}

\textbf{Motivation -- Impact of Tuning:}
In Fig.~\ref{fig:power_profiles}, we show the power profiles for a regular advanced encryption standard (AES) core
when operating the core under different driver strengths and VCCs.
We observe that dynamic tuning---that is, runtime reconfiguration of switching drivers and/or utilized
VCCs---results in profiles that largely
overlap with the baseline profiles.

{Such interspersion of power profiles as shown in Fig.~\ref{fig:power_profiles} represents a major challenge for PSC
attacks} as follows. Assuming the same
text is processed, using the same key, but under varying tuning settings, a large number of different power values can arise within the
resulting power profiles. The reverse also applies: for the same power value observed, a large range of different
possible texts and/or different possible keys may be underlining of the crypto computation.
Naturally, such ambiguity can be quite misleading for analytical models that are at the heart of PSC attacks.

\textbf{Contributions:}
{As indicated, prior art did overlook the potential of jointly tuning driver strengths and VCCs in general,
let alone for dynamic runtime modes.} In this work, we
address this gap.
We build up a multi-part methodology to thoroughly study various 
scenarios for joint tuning of driver strengths and VCCs, applicable for ASIC as well as FPGA fabrics.

We take the following steps:
\begin{enumerate}
\item We develop a simple circuit-level scheme for tuning, which is applicable for ASIC as well as FPGA fabrics.
\item We implement a CAD flow for design-time evaluation of ASIC power profiles at runtime,
enabling proper security assessment of ICs before tape-out.
\item We implement a correlation power analysis (CPA) framework for a thorough and comparative security analysis of tuning.
\item As key contribution of this work, we conduct an extensive experimental study of a regular AES design, implemented in ASIC as
well as FPGA fabrics, under various tuning scenarios.
\item Finally, we derive design guidelines for secure and efficient joint tuning in ASIC and FPGA fabrics.
\end{enumerate}
We emphasize that our work is not meant to compete with or replace prior art for PSC 
countermeasures, but rather to extend the landscape of available options at a foundational level. Joint tuning
can be either used as stand-alone measure, as it is done in this work at hand, or to further complement prior countermeasures.

\textbf{Findings:}
In our experimental study, we observe the following.
\begin{enumerate}

\item For the ASIC design based on a GlobalFoundries 55nm technology,
we find that a)~static design-time tuning
may improve the resilience, but only to limited degrees, and can sometimes even counteract it.
In contrast,
b)~dynamic runtime tuning always renders the AES core more resilient, namely up to 235\% as resilient as the untuned
baseline.\footnote{%
	This quantitative finding is conservative, as it is based on design-time evaluation
	without impact of layout effects, let alone measurement noises. Thus,
	for attacks on real hardware, we can assume a larger impact---we confirm this via FPGA implementation.}

\item We confirm our key finding---that dynamic runtime tuning is more effective---in the field, using the Sakura-X
FPGA board based on a 28nm technology.
Here, the AES core is rendered >11.8x (i.e., at least 11.8 times) as resilient.

\item
Regarding trade-offs for resilience versus layout overheads, we find them
reasonable with, e.g., $\approx$10\% impact on critical-path delay for the
most resilient FPGA tuning scenario.

\end{enumerate}

\textbf{Release:}
We will release source codes for our methodology, as well as empirical artifacts, post peer-review in~\cite{anon_web}.

\section{Background}
\label{sec:bg}

\subsection{Side-Channel Attacks}
\label{sec:bg:SCA}

Side-channel attacks infer sensitive information by observing and analysing physical channels
established by ICs during operation~\cite{zhou05}. These channels are leaking some kind of information due to the
basic workings of the underlying circuitry, but also due to micro-architectural implementation decisions. For the
latter, e.g.,
timing behavior and speculative execution in modern processors has been demonstrated as vulnerability~\cite{osvik05}.
For the classical PSC, e.g.,
it is well-known that the secret key for AES
can be inferred by analysing the data-dependent power
consumption~\cite{brier04,skorobogatov12}.
	
Different PSC attacks have been demonstrated, like
correlation power analysis
(CPA)~\cite{brier04}, mutual information analysis~\cite{gierlichs2008mutual}, 
or machine learning-based techniques~\cite{picek2017side}.
Furthermore, there are more generic, analytical
approaches
	like test vector
leakage assessment (TVLA)~\cite{schneider15},
	architecture correlation~\cite{yao20},
etc.

Without loss of generality (w/o.l.o.g.), we focus on the CPA attack in this work.
CPA is well-established and used widely throughout the literature.
CPA is effective, e.g.,
CPA requires on average fewer traces than DPA~\cite{brier04,massimo08,Fei2015}.
More details are explained in Sec.~\ref{sec:method:CPA_framework}.

\subsection{Prior Art for Countermeasures}
\label{sec:bg:prior}

Various countermeasures against PSC attacks {have} been proposed over the years,
including masking and hiding~\cite{li17_SCA%,9015513
}, voltage
switching~\cite{8361769,9006696%,10.1145/2345770.2345774%,DBLP:journals/corr/abs-1907-09440
}, noise injection~\cite{bellizia18%,8351968
}, etc.
Essentially, these countermeasures seek to de-correlate the observable
power consumption from the sensitive crypto operations.

More specifically, masking and hiding approaches do restructure and reimplement the design such that
sensitive operations are decomposed/split at the functional as well as the circuit level.
However, the overheads for such schemes can scale quadratically with the related
security requirements, making efficient implementations quite challenging~\cite{10.1007/978-3-319-52153-4_6}.
Voltage switching can be enabled by, e.g., multiple voltage domains and related control circuitry, or by integrated voltage
regulators (IVRs)~\cite{8361769}.
Note that IVRs are commonly available in modern IC designs, as they can enable significant power savings.
Noise injection, like interposing random data into
redundantly designed register paths~\cite{bellizia18}, is effective but can also {incur} considerable area and power costs.

Driver strengths, while directly impacting power profiles and thus the resilience against PSC attacks, 
have been largely overlooked in prior art. To the best of our knowledge, ``Karna''~\cite{karna}
is the only recent work that has explicitly studied the role of driver strengths (known as gate sizes in~\cite{karna}), along with
VCCs and threshold voltages. The authors found that varying strengths has also varying impact on the resilience.
Tuning of these parameters, however, was limited to static design-time tuning for ASICs.\footnote{%
In contrast, we study both static and dynamic tuning, for ASIC as well as FPGA fabrics. We find that static tuning
has limited effects; this is in agreement with findings in ``Karna.'' We also find that static
tuning is even counteractive for FPGA implementation, whereas ``Karna'' did only study ASIC implementation.}

\section{Threat Model}
\label{sec:tm}

We consider a stringent threat model for PSC attacks as follows.

Attackers can only act as passive observers.
That is, attackers do have direct/indirect access to the ASIC or FPGA, but only for monitoring the power
consumption and the cipher-texts. Attackers have no control of plain-texts and no control over the power supply.

We assume that attackers are fully aware of our countermeasure's working principle.
However, given that operation of the tuning implementation (Sec.~\ref{sec:method:tuning})
is randomized, randomly switching between different tuning scenarios, and given that power profiles for
different tuning scenarios are considerably interspersed (recall Fig.~\ref{fig:power_profiles}),
attackers cannot ascertain the specific driver strength and VCC underlying for any particular point in
time or operation. Accordingly, attackers cannot explicitly separate the multiple distributions underlying the power
profiles, which will hinder any analytical attack model.

\section{Methodology}

\subsection{Runtime Tuning of Driver Strengths, VCCs}
\label{sec:method:tuning}

This section outlines implementation options for the key idea of our work, that is dynamic tuning of driver strengths
and VCCs. Other options could be devised as well, e.g., toward a more optimized, cell-level integration of
different driver strengths.

Registers in general, and those holding AES texts in particular, are most
relevant for PSC attacks, since they build up considerable correlation between the processed data and the observable
power consumption;
see also
 Sec.~\ref{sec:method:CAD_flow} and Sec.~\ref{sec:method:CPA_framework}.
Thus, for both ASIC and FPGA implementations, we focus on registers.

\textbf{Implementation in ASICs:}
For static driver-strength tuning during design time, we simply reconfigure the strength for each register
of choice. We randomly select, w/o.l.o.g., either the lowest or highest available strength. To maintain the selected
strengths throughout the design flow, we mark these register as ``don't touch.''

\begin{figure}[tb]
\includegraphics[width=0.90\columnwidth%, height=5.5cm%, keepaspectratio
]{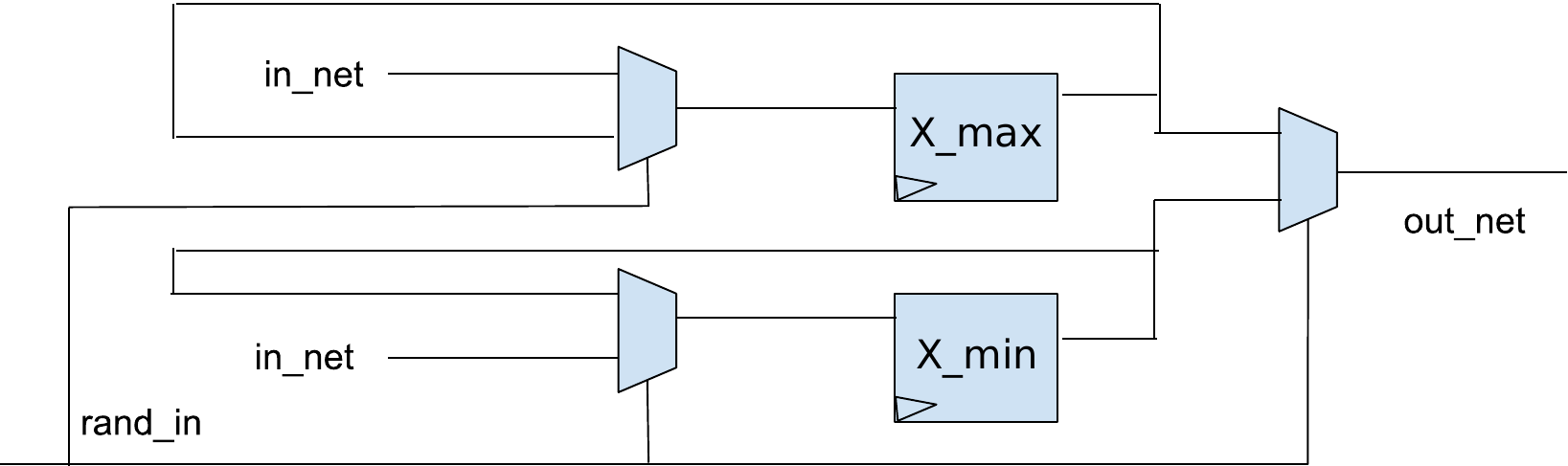}
\littlesmallerspace
	\caption{Implementation principle of dynamic tuning of driver strengths.}
	\label{fig:tune_imp}
\smallerspace
\smallerspace
\end{figure}

For dynamic tuning of driver strengths, we implement tunable registers
as outlined in Fig.~\ref{fig:tune_imp} and described next.
For each register of choice, we replace it with a pair of registers, again w/o.l.o.g.~one with lowest and one with
highest available strengths, respectively.
Register pairs are marked as ``don't touch''
such that the strengths are maintained.
We reconnect the original register's nets through two additional multiplexers
(MUXes) such that only one of the two registers is randomly selected for operation at a time; the other register is
feeding back itself the current data, i.e., is guaranteed to not toggle at that point in time.

For design-time evaluation, VCC tuning is mimicked through cell and library configurations.
For actual ASIC implementations, we assume IVRs or other tuning features to be
available.
Note that
requirements for such
would be reasonable; dynamic VCC tuning is required only once per
full AES round, not every clock cycle. This is because the CPA attack focuses on the last (or first) intermediate
round~\cite{brier04}; other attacks follow similar principles of attacking specific parts of the AES operation.

\textbf{Implementation in FPGAs:}
Note that common FPGA fabrics do not provide the option for reconfiguring cell driver strengths.
However, IO pins can be reconfigured for driver
strengths and other parameters.
Thus, we implement tuning on FPGAs as follows.

For each register of choice, we additionally connect its output with two IO pins, again w/o.l.o.g.~one with lowest and one with highest
available IO driver strengths, respectively.
Similar to the ASIC implementation, we use additional MUXes to randomly select only one IO pin to be driven at
a time.

For static tuning scenarios, we simplify the above implementation by additionally connecting and hard-wiring each register of choice to only one
IO pin of lowest/highest IO driver strength.

For VCC tuning, we assume some tuning features are available, like FPGA on-board voltage regulators.

\subsection{CAD Flow for Design-Time Evaluation of ASIC Power Profiles}
\label{sec:method:CAD_flow}

The CAD flow described here serves to investigate the role that joint tuning of driver strengths and VCCs plays
against PSC attacks early on, in a design-time simulation environment, without need for FPGA implementation
or even IC tape-out and measurements.\footnote{%
We still conduct FPGA implementation, measurements,
and related analysis later on, to verify our findings for practical attack scenarios and across hardware
fabrics.}
Note that the idea of such CAD-flow-based investigation is not new; similar approaches
have been taken in, e.g., \cite{karna,Sadhukhan2019}. Still, the flow described here has been devised
     independently of prior art, and also verified within other studies
[omitted for blind review].

\begin{figure}[tb]
\includegraphics[width=.95\columnwidth]{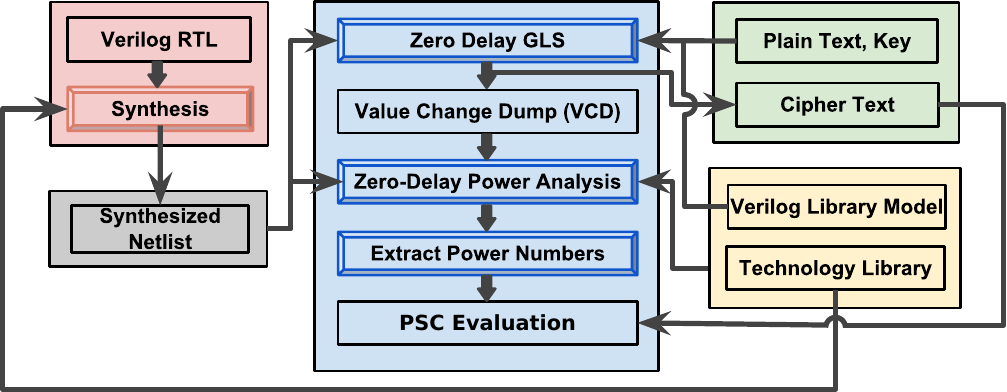}
\littlesmallerspace
	\caption{CAD flow for design-time evaluation of zero-delay power profiles of ASICs.}
	\label{fig:CAD_flow}
\smallerspace
\smallerspace
\end{figure}

Our flow (Fig.~\ref{fig:CAD_flow}) takes as inputs: (i) the register-transfer level (RTL) code of the design to be
evaluated, e.g., a regular AES crypto core, (ii) the
standard-cell library of choice, and (iii) sets of plain-texts and keys. The latter can also be randomly generated
within the flow itself.
The flow provides the zero-delay power values,
i.e. power values without any impact of layout effects or noises, for the design's circuit-level computation as
triggered by the plain-text and key inputs.
These power values are then utilized for PSC evaluation/security analysis (Sec.~\ref{sec:method:CPA_framework}).

Next, we describe the flow in some more detail.
For the implementation, we employ regular commercial tools (Sec.~\ref{sec:setup:tools}).
We will release source codes for our CAD flow post peer-review in~\cite{anon_web}.

\textbf{Step 1:}
We synthesize the AES core's RTL. We verify the functionality of the obtained gate-level netlist
using a Verilog testbench, with randomized sets of plain-texts and keys.
We also confirm the functionality of the design, using software simulation of the crypto operations and cross-checking of
the two sets of cipher-texts.

\textbf{Step 2:}
We perform zero-delay gate-level simulation of the design, to generate a value change dump (VCD) file.
Note that VCD files are
well-established for simulation purposes.

\textbf{Step 3:}
The VCD file is then used for power simulation of the synthesized gate-level netlist.
To limit simulation efforts---without comprising the accuracy for the PSC attack evaluation---we focus only on the relevant
time intervals, i.e., the last (or first round) of AES, which are the ones sufficient to attack~\cite{brier04}.

\textbf{Scope of Simulations:}
Instead of performing full-scale transient simulations, which would also capture noises induced by glitching activities, here we leverage
noise-free, zero-delay simulations.
For our notion of tuning driver strengths and VCCs, glitches are less relevant; tuning has significant impact
on power profiles overall (recall Fig.~\ref{fig:power_profiles}), not only on
glitching activities.

For such zero-delay simulation, all power-consuming transitions occur simultaneously for
the clock edge. Thus, peak-power values, which are of particular relevance for PSC attacks, can be easily extracted.
Further, note that register are generally contributing the largest shares of dynamic power consumption.
While the registers holding the secret key itself are not switching, thus not providing any leverage for PSC attacks,
other registers do switch. In fact, those register that are holding the texts of intermediate AES rounds
incur considerable switching activities
      by design, due to the confusion and diffusion properties of the AES crypto algorithm, and can thus be well
      correlated against.

\subsection{CPA Framework for Security Analysis}
\label{sec:method:CPA_framework}

The CPA framework described here serves for an empirical security analysis, yet in a thorough manner and backed by
solid analytical formalism.
We will release source codes
post peer-review in~\cite{anon_web}.

As indicated, CPA is known to be effective~\cite{brier04,massimo08,Fei2015}.
At the heart of CPA is the linear {Pearson correlation coefficient (PCC)}, used to quantify the
relationship between actual power profiles and hypothetical power profiles.
The latter are typically built up by enumeration of all
byte-wise possible keys~\cite{brier04}.
After building up correlation over a number of traces---obtained in any way, e.g., via design-time power
simulations using the above CAD flow or via measurements---the most promising
candidates for all bytes are concatenated to form the guess of the correct key.

We take the following steps in our framework.

\textbf{Step 1:}
Note that, since registers consume a significant share of
dynamic power during signal transitions, the Hamming distance (HD) for the registers'
data before and after switching operations is established as simple, yet effective, \textit{HD power model}~\cite{brier04}.

Now, as indicated, sets of hypothetical power values are to be derived for all possible key values.
This is done using the HD power model, namely by reverting the AES last-round operation using the observed cipher-texts, and
computing and memorizing the HD when considering all possible key values for that reverse operation.

\textbf{Step 2:}
Using the PCC formalism, the actual power traces---again, can obtained in any way---are
correlated against all hypothetical power profiles. The profile resulting in the
highest PCC value across a number of traces is assumed to represent the correct key.
As indicated, the correlation analysis can be conducted at the byte level (instead of bit level)~\cite{brier04},
which is essential to manage complexity for exploring
the search space of all possible keys.

Instead of considering all available traces at once for this correlation analysis, we thoroughly and step-wise explore
the range of how many traces are needed at least until disclosure of the correct key with certain
confidence. See Sec.~\ref{sec:metrics:sec} for more details.

\textbf{Attack Versus Security Analysis:}
Acting as designers, we can readily verify the key guess for any CPA run during the security analysis.
An attacker, however, has to monitor the progression of PCC values for all the possible key hypotheses throughout a more or less
large number of traces; only once the best candidate shows a significant PCC outlier among all other candidates,
can the attacker assume to have successfully inferred the correct key.

We take the attacker's approach for parts of our study as well, namely for realistic pre-processing (i.e.,
without relying on the actual correct key) of noisy power traces obtained for FPGA measurements. There, any sub-set of
traces that does not exhibit a sufficiently significant PCC outlier is rejected as too noisy.

\section{Empirical Study: Setup}
\label{sec:setup}

\subsection{Experimental Setup}

\begin{figure}[tb]
\includegraphics[width=0.70\columnwidth, height=4.4cm%, keepaspectratio
]{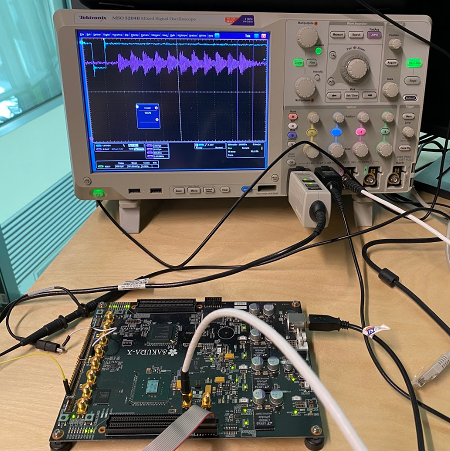}
\littlesmallerspace
	\caption{FPGA measurement setup.}
	\label{fig:lab_setup}
\smallerspace
\smallerspace
\end{figure}

\begin{figure*}[tb]
\includegraphics[width=\textwidth]{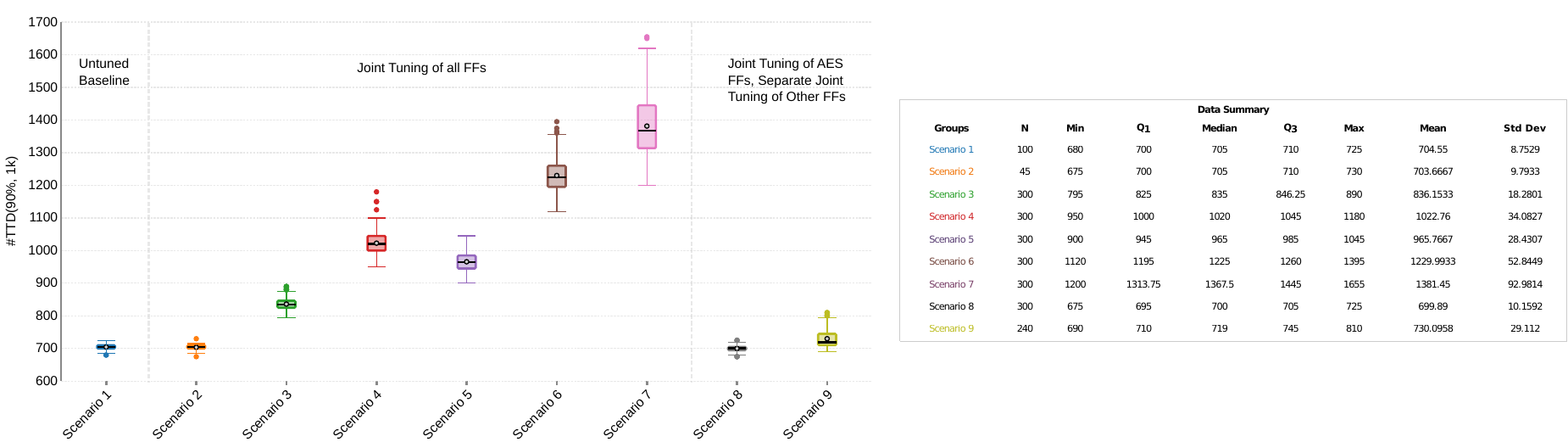}
	\smallerspace
	\smallerspace
	\caption{CPA results for the ASIC implementation. See the main text for description of the different
		scenarios. Also recall that, for each data point underlying each box,
			there is a robust and thorough sampling process underlying (Sec.~\ref{sec:metrics:sec};
					Footnote~\ref{fn:sampling}).}
	\label{fig:ASIC_CPA}
	\littlesmallerspace
\end{figure*}

\textbf{Tools:}
\label{sec:setup:tools}
We devise the CAD flow and perform ASIC implementation using standard commercial tools, i.e., Synopsys DC for logic
synthesis and Synopsys VCS for gate-level power simulation. We devise custom tcl scripts for the CAD flow integration
and bash scripts for data management and processing.
We use Xilinx ISE Webpack suite for FPGA implementation.
We implement the CPA framework in C++, based on the release in~\cite{CPA_yunsi}.

\textbf{Design:}
We utilize a regular AES core, with 128-bit keys and 128-bit texts processed in electronic code book (ECB) mode.
We release the RTL post peer-review in~\cite{anon_web}.

\textbf{Implementations:}
For the ASIC implementation, we employ a commercial 55nm technology by GlobalFoundries, for logic synthesis and
zero-delay, gate-level power simulation.
For the FPGA implementation, we use a Sakura-X board, specifically its Kintex-7 FPGA chip, which is manufactured
in a 28nm technology. We build up a common FPGA measurement setup (Fig.~\ref{fig:lab_setup}).
We tune VCCs using the FPGA's on-board core-voltage regulator.

Naturally, the ASIC and FPGA implementations differ considerably in terms of (i) available driver strengths and VCCs,
(ii) technology nodes and hardware fabrics, and (iii) noise profiles.
Such diversity is essential to confirm and generalize our findings.

\textbf{Metrics and Workflow for Security Analysis:}
\label{sec:metrics:sec}
We report the minimal number of traces needed to disclosure 
as \T{c}{t}, i.e., for a confidence value $c$ across $t$ randomized CPA trials.

Throughout all experiments, we report \T{90}{1k} which means that $\geq$900 out of 1,000 randomized CPA trials succeed
for the reported number of traces.
To determine \T{90}{1k} values accurately, we conduct multiple CPA campaigns as follows.
Each campaign is run independently in steps, where an increasing number of randomly selected plaint-texts and corresponding
power traces are made available to the CPA framework. More specifically, for each campaign step, 1,000 randomized CPA
trials are conducted on 1,000 different sets of randomly selected texts and corresponding traces.
The success rate is tracked and more and more steps are taken, until the point of 90\% confidence is reached, i.e.,
900/1,000 trials succeed.
While such workflow is computationally intensive, it is trivial to parallelize, and essential for a robust security analysis.

For the exploration of different tuning scenarios,
we repeatedly conduct CPA campaigns with many trials, as outlined above, all while
maintaining the overall sets/pool of plain-texts and keys. Doing so is
important for fair comparison across tuning scenarios.

We consider sets of 5k traces for the ASIC implementation and 15k--170k traces for the FPGA implementation.\footnote{%
For the FPGA implementation, we observe that 15k traces are sufficient for breaking less resilient scenarios, whereas around 200k traces
are still insufficient for breaking the more resilient scenarios. As indicated, we pre-process measurement traces,
similar to what an attacker would do, to
reject noisy traces. After measuring 200k traces, we split them into 20 by 10k sets, and had to reject three sets;
thus, 170k traces remain.}
   For each step in any CPA campaign, we increase the number of available traces by 5 and by 15,
   respectively, for the ASIC and the FPGA implementation.

\textbf{Metrics and Workflow for Layout Analysis:}
We report power, performance, and area (PPA) numbers for ASIC and FPGA implementations.
For ASIC implementation, performance and area is reported from logic synthesis using DC.
Power
is reported as average peak power, derived from the same gate-level simulations used for security analysis.
For FPGA implementation, performance and area---the latter in terms of utilization of flip-flops (FFs) and look-up
tables (LUTs)---are reported from ISE runs.
Power is reported as average peak power from measurements.
We also report additional IO pins used for implementing tuning in FPGA.

\subsection{Tuning Settings}
We consider the following tuning settings for our study.
Each setting comprises different scenarios, in terms of static versus dynamic tuning
and in terms of driver strengths, VCCs available for tuning in the ASIC versus FPGA
implementation.
\begin{enumerate}

\item[(I)] All FFs are tuned to the same driver strength and VCC.

	\begin{enumerate}

	\item[(Ia)] Static tuning only.

	\item[(Ib)] Static and dynamic tuning, covering all combinations of static/dynamic tuning for driver strengths and VCCs.

	\item[(Ic)] Dynamic tuning only.

	\end{enumerate}

\item[(II)] FFs holding AES texts versus all other FFs are separated into two groups.
Groups of FFs are tuned differently, whereas all FFs within a group are tuned the same.
This setting is motivated by the potential need for a more limited and less costly implementation; 
see also Sec.~\ref{sec:experiments:layout}.
	\begin{enumerate}

	\item[(IIa)] Static tuning only.

	\item[(IIb)] Dynamic tuning of FFs holding AES texts; static tuning of all other FFs.

	\end{enumerate}

\end{enumerate}
Note that, for Setting (II), we refrain from considering further possible scenarios, like dynamic tuning of FFs holding AES
texts along with different dynamic tuning of all other FFs.
This is w/o.l.o.g., due to practical limitations on the number of available IO pins for dynamic tuning on our FPGA
of choice.

\section{Empirical Study: Security Analysis}
\label{sec:experiments:security}

\subsection{ASIC Implementation}

Detailed results are provided in Fig.~\ref{fig:ASIC_CPA}; related observations are presented next.
Note that the numbering of scenarios below matches that in Fig.~\ref{fig:ASIC_CPA}.
Also note that, in Sec.~\ref{sec:experiments:security:summary}, we streamline and summarize
findings for both ASIC and FPGA implementations.

\begin{enumerate}

    \item \textit{Untuned Baseline:}
The regular AES design, without any tuning. VCC is set to 1.08V for all FFs and all other gates.
Driver strengths are set automatically by logic synthesis.
Here, 100 CPA campaigns are conducted,
resulting in $N=100$ data points for the \T{90}{1k} metric (Sec.~\ref{sec:metrics:sec}), but recall that
many more CPA trials are underlying for a robust analysis.\footnote{%
	\label{fn:sampling}
As indicated, each campaign progresses in steps of
more traces becoming available (w/o.l.o.g., 5 traces for the ASIC implementation), and for each step 1,000 trials are run.
To avoid running many trials with too few
traces to begin with, we initially conduct some exploratory sampling, to determine a reasonable starting point for all
campaigns; we found 600 traces suitable for this scenario. Thus, there are 100 campaigns run, with 16--25 steps per
campaign (covering the observed min and max points of 680 and 725 for \T{90}{1k}, respectively), with 1,000 trials per step, resulting in 1.6--2.5 million CPA trials in total, just for exploring this tuning
scenario. All other scenarios are explored in the same thorough manner.}

\end{enumerate}

\textit{Tuning Setting (I):} For static or dynamic tuning of all FFs,
	we consider the following scenarios.

\begin{enumerate}
\setcounter{enumi}{1}

    \item \textit{Static X, VCC:}
    All combinations for all five available driver strengths, ranging from X0.5 to X4, as well as three
    available VCCs, ranging from 0.9V to 1.08V, are considered, resulting in 15 tuning configurations and, across
    three CPA campaigns, in $N=45$ data points.

    This is the least resilient scenario across Setting (I), with little difference to the untuned baseline.
    Along with low standard deviation (SD) across all 45 combinations, this indicates that
    \ul{exclusively static tuning is not effective}.

    \item \textit{Static X0.5, Dynamic VCC:}
    Here, 100 runs for randomized, dynamic VCC tuning across 0.9--1.08V, are considered in three CPA campaigns,
    resulting in $N=300$ data points.

    This scenario is more resilient than static tuning (2), but
    less resilient than dynamic VCC tuning for higher driver strengths (4), and also less resilient than
    driver-strength tuning for static VCCs (5), (6).
    This indicates that dynamic tuning can be beneficial, when applied thoughtfully.

    \item \textit{Static X4, Dynamic VCC:}
    Same setup as in (3), except driver strengths are set to X4 for all FFs.

    As indicated, this scenario is more resilient than VCC tuning for lower driver strength (3). It is also somewhat
    more resilient than driver-strength tuning for low, static VCC (5). These observations, together with those for
    (2), imply that \ul{high driver strengths
	    can be beneficial}
	    and \ul{dynamic VCC
	    tuning can be beneficial}.

    \item \textit{Dynamic X, Static 0.9V:}
    Here, 100 runs for randomized, dynamic driver-strength tuning across X0.5--X4, are considered in three CPA campaigns, resulting in
		    $N=300$ data points.

    As indicated, this scenario is on average more resilient than VCC tuning for low driver strength (3), but
    somewhat less resilient than VCC tuning for high driver strength (4).

    \item \textit{Dynamic X, Static 1.08V:}
    Same setup as in (5), except VCCs are set to 1.08V for all FFs.

    On average, this scenario is more resilient than all prior ones. Considered along with (5), this
    implies that, \ul{for dynamic driver-strength tuning, high VCCs can be beneficial}.

    \item \textit{Dynamic X, Dynamic VCC:}
    Here, 100 runs for randomized, dynamic driver-strength tuning across X0.5--X4 and dynamic VCC tuning across 0.9--1.08V,
    are considered in three CPA campaigns, resulting in
		    $N=300$ data points.

    This is the most resilient scenario. This implies that \ul{dynamic tuning of both driver strengths
    and VCCs provides superior resilience}, namely on average 196\% and up to 235\% as resilient as both static tuning and the
	    untuned baseline.

\end{enumerate}

For \textit{Tuning Setting (II)}, separate tuning of FFs holding AES texts versus all other FFs, we consider the following scenarios.

\begin{enumerate}
\setcounter{enumi}{7}

    \item \textit{Static Only:}
    For each of the two groups, all combinations of five driver strengths, ranging from X0.5 to X4, as well as two VCCs, 
    0.9V and 1.08V, are considered, resulting in 100 configurations, $N=300$ data points for three CPA campaigns.

    Here, we observe a similarly low resilience as with static-tuning Scenario (2), again with low SD across all
    different configurations.
    This re-iterates that \ul{static tuning alone is not effective, not even in any particular tuning configuration}.

    \item \textit{Static and Dynamic:}
    All combinations of two corner-case driver strengths, X0.5 and X4, and VCCs, 0.9V and 1.08V, are
    considered for static tuning of other FFs. At the same time, all six possible, non-redundant combinations for dynamic
    tuning using different configurations for driver strengths and VCCs are considered for FFs holding AES texts.
    In other words, this scenario explores dynamic tuning of FFs holding AES texts, while statically tuning all
    other FFs.
    All combinations are explored via ten CPA campaigns, resulting in
    $N=4*6*10=
    240$ data points.

    Also here, we observe a low resilience, similar to
    (2) and (8), albeit with
    a higher SD, implying that some particular configurations are more promising.
    Still, \ul{limiting dynamic tuning to only the FFs holding AES texts is not effective}.\footnote{While this
	    applies for the ASIC implementation, we observe that, for the FPGA implementation, such limited tuning can
		    still be effective.}

\end{enumerate}

\subsection{FPGA Implementation}

Next, we elaborate on our findings for the FPGA implementation.
In Sec.~\ref{sec:experiments:security:summary}, we streamline and summarize all
findings.

Given that available driver strengths and VCCs, as well as noise profiles, differ vastly from those of the ASIC
implementation, quantitative results cannot be compared across these implementations.
More importantly, however, observations are verified across hardware fabrics and even
technology nodes, rendering our insights on the prospects of tuning robust.

\begin{figure*}[tb]
\includegraphics[width=.90\textwidth]{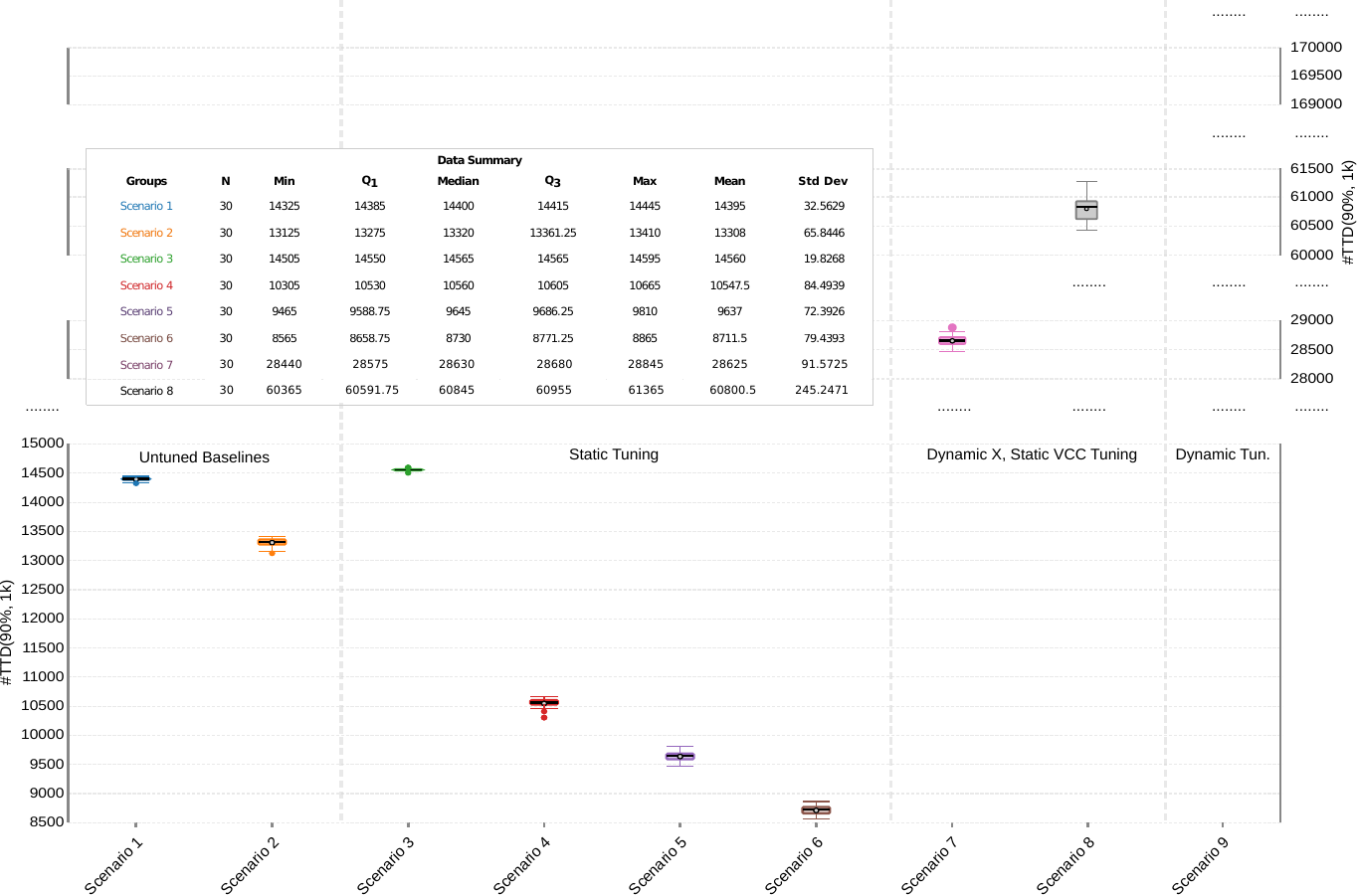}
\littlesmallerspace
	\caption{CPA results for the FPGA implementation. See the main text for description of the different
		scenarios.
			Note the varying ranges for \T{90}{1k} across the different tuning scenarios.
			Also recall that, for each data point underlying each box,
			there is a robust and thorough sampling process underlying (Sec.~\ref{sec:metrics:sec};
					Footnote~\ref{fn:sampling}).}
	\label{fig:FPGA_TTD}
\littlesmallerspace
\end{figure*}

Detailed results
are provided in Fig.~\ref{fig:FPGA_TTD}; related observations are discussed next.
The scenario numbering matches that in Fig.~\ref{fig:FPGA_TTD}.

\begin{enumerate}

	\item \textit{Untuned Baseline, 0.955V:}
	The regular AES design, without any tuning. VCC is set to 0.955V for all FFs.

	\item \textit{Untuned Baseline, 1.055V:}
	The regular AES design, without any tuning. VCC is set to 1.055V for all FFs.
This scenario is less resilient then (1), suggesting that
	\ul{lower VCCs \textit{can} be beneficial}, especially as long as driver strengths are not tuned dynamically; see
		also the remaining scenarios.

\end{enumerate}

Recall that we use additional IO pins for
driver-strength tuning in the FPGA implementation (Sec.~\ref{sec:method:tuning}).
Since IO pins are limited, we also want/need to limit the number of FFs that are tuned.
Thus, counterparts for promising configurations observed for the ASIC implementation, namely Setting (I) in general and the most resilient Scenario (7) in
particular---dynamic tuning of both driver strengths and VCCs for all FFs---are impractical for the FPGA implementation.

Accordingly, we skip directly to \textit{Tuning Setting (II)}, separate tuning of FFs holding AES texts versus all other FFs.
Again considering limited numbers of IO pins, we do not specifically tune other FFs here, but only FFs
	holding AES texts.
	We consider the following resulting scenarios.
   30 CPA campaigns are conducted
(i.e., $N=30$ data points)
	for each scenario unless stated otherwise.

\begin{enumerate}
\setcounter{enumi}{2}

    \item \textit{Static X4, 0.955V:}
This scenario is only slightly more resilient than the untuned baseline for the same VCC, indicating that static
tuning with low driver strength is not effective.

    \item \textit{Static X4, 1.055V:}
Here, there is a significant \textit{drop} in resilience.
Considering together with (3) and the remaining scenarios for static tuning, i.e., (5) and (6), this indicates that
\ul{static tuning is most often counterproductive}, due to increased information leakage
incurred via toggling IO pins according to AES texts held in the related FFs.

    \item \textit{Static X16, 0.955V}

    \item \textit{Static X16, 1.055V}

    \item \textit{Dynamic X, Static 0.955V:}
This scenario represents a strong turning point.
On average, when compared to the corresponding untuned and statically-tuned baselines,
resilience is increased to 199\% and 197--297\%, respectively.
This indicates that \ul{dynamic driver-strength tuning is effective}.
Unlike with static tuning, where any related changes in power profiles can still be well correlated against, such dynamic tuning
significantly interrupts the correlation working principle, by interspersing power
profiles \textit{in a randomized manner} such that different texts may well be related to the same power values and vice versa, as motivated in
Sec.~\ref{sec:intro}.

    \item \textit{Dynamic X, Static 1.055V:}
This scenario represents another turning point, as in high a VCC notably improving resilience again, unlike for the untuned
baselines or static-tuning scenarios. This implies that \ul{high VCCs can be beneficial in combination with
dynamic driver-strength tuning}. 

    \item \textit{Dynamic X, Dynamic VCC:}
This is the most resilient scenario by far.
Here, we consider even 100 CPA campaigns for robust sampling, and find that
none can break the tuning-induced resilience, even when considering all 170k available traces at once.\footnote{%
	Accordingly, there are no corresponding \T{90}{1k} data points included in Fig.~\ref{fig:FPGA_TTD}.}
This implies that resilience is increased by \textit{at least} 11.8x over the untuned baseline (1).
This clearly shows that \ul{joint and dynamic tuning is by far most resilient}.

\end{enumerate}

\subsection{Summary}
\label{sec:experiments:security:summary}

Our findings for both the ASIC and FPGA implementations are:
\begin{enumerate}
\item Dynamic VCC tuning is promising, but limited on its own;
\item Dynamic driver-strength tuning, along with high VCCs or dynamic VCC tuning, is most
effective;
\item Tuning of all FFs is promising, but is also limited in practice (by available IO pins for the FPGA implementation and
by overheads for the ASIC implementation; see Table~\ref{tab:ASIC_PPA} below);
\item Static tuning is least effective in general and even counterproductive for the FPGA implementation (where implicit masking by
environmental noises can be nullified when using high VCCs and/or high driver strengths for tuning).
\end{enumerate}

\section{Empirical Study: Layout Analysis}
\label{sec:experiments:layout}

\begin{table}
\setlength\tabcolsep{2.0pt}
	\small
	\caption{Layout Analysis for ASIC Implementations}
	\smallerspace
	\label{tab:ASIC_PPA}
	\begin{tabular}{cccc}
		\toprule
		\multirow{3}{*}{Design}& Avg. Peak & Critical-Path & Std.-Cell \\
		& Power [mW]
		& Delay [ns] & Area [$\mu m^2$] \\
		& 0.9V / 1.08V & 0.9V / 1.08V & 0.9V / 1.08V \\
		\toprule
		Baseline & 2.709 / 3.100 & 9.64 / 9.79 & 54,928 / 43,639 \\
		\midrule
		All FFs & 3.134 / 3.764 & 14.13 / 11.86 & 67,873 / 57,300 \\
		Tunable & (+15.69\% / +21.42\%) & (+46.68\% / +21.14\%) & (+23.57\% / +31.30\%) \\
		\midrule
		AES-Text & 2.779 / 3.237 & 14.13 / 11.68 & 57,272 / 46,324 \\
		FFs Tunable & (+02.58\% / +04.42\%) & (+46.68\% / +19.31\%) & (+04.27\% / +06.15\%) \\
		\bottomrule
	\end{tabular}%\\[1mm] % only needed if table footnote is used
\littlesmallerspace
\end{table}

\textbf{ASIC Implementation:}
See Table~\ref{tab:ASIC_PPA}.
Naturally, layout costs are larger when all FFs are tunable, whereas costs are reasonable if only FFs holding
AES texts are tunable.

We argue that costs may well be amortized for large-scale ASIC designs with many modules. In
contrast, to study upper limits of overheads, here we consider a stand-alone AES core.
Besides, some optimized cell-level integration of different driver strengths and tuning peripherals may be attainable
in future work.

\begin{table}
	\small
	\caption{Layout Analysis for FPGA Implementations}
	\smallerspace
	\label{tab:FPGA_PPA}
	\begin{tabular}{cccc}
		\toprule
		\multirow{3}{*}{Design}& Avg. Peak & \multirow{2}{*}{Critical-Path} & \multirow{2}{*}{FFs / LUTs} \\
		& Power [mW]
		& \multirow{2}{*}{Delay [ns]} & \multirow{2}{*}{Util. [\# / \#]} \\
		& 0.9V / 1.08V & & \\
		\toprule
		Baseline & 0.969357 / 1.07049 & 9.052 & 952 / 3,137 \\
		\midrule
		\multirow{2}{*}{Static X4} & 0.972771 / 1.07352 & 10.352 & 965 / 3,118 \\
		& (+00.35\% / +00.28\%) & (+14.36\%) & (+01.37\% / -00.61\%) \\
		\midrule
		\multirow{2}{*}{Static X16} & 0.971117 / 1.07347 & 10.352 & 965 / 3,118 \\
		& (+00.18\% / +00.27\%) & (+14.36\%) & (+01.37\% / -00.61\%) \\
		\midrule
		\multirow{2}{*}{Dynamic} & 0.973676 / 1.07214 & 9.994 & 1,028 / 3,183 \\
		& (+00.45\% / +00.15\%) & (+10.41\%) & (+07.98\% / +01.47\%) \\
		\bottomrule
	\end{tabular}%\\[1mm] % only needed if table footnote is used
\littlesmallerspace
\end{table}

\textbf{FPGA Implementation:}
See Table~\ref{tab:FPGA_PPA}.
Note that, for both static-tuning designs, critical-path delays and utilization are the same; this is because only the
driver-strength configurations for the additional IO pins differs here, whereas the core circuitry remains the same.
Overall, we observe marginal impact on power as well as utilization,
along with some overheads for
delays.

Delay overheads are due to large-scale changes imposed on placement and routing,
after connecting the circuitry to the additional IO pins used for tuning, as we have observed via ISE PlanAhead.
An iterative design strategy
might reduce overheads;\footnote{%
	For example, the strategy could be (i) placement and
routing, (ii) selection of nearby IO pins for assignment to tuned FFs, (iii) evaluation of layout overheads, and (iv)
	repeated selection/assignment of IO pins, guided by worst-case timing overheads, etc.}
	currently, we assign from available IO pins arbitrarily to FFs to be tuned.

\section{Design Guidelines for Tuning}

Recall the key findings for the security analysis, summarized in Sec.~\ref{sec:experiments:security:summary}.
Considering these together with the layout analysis in Sec.~\ref{sec:experiments:layout}, we propose the following design
guidelines.

Static tuning is discouraged.
Dynamic and joint tuning should be applied whenever possible. Otherwise, dynamic tuning of driver strengths is
preferred as simple, yet effective, alternative. This is because (i)~VCC tuning requires some IVR or other
tuning features, whereas driver-strength tuning can be implemented at circuit level at its own,
and (ii)~dynamic driver-strength tuning is more resilient.

Tuning of all FFs can be considered when the relatively high layout overheads for an ASIC implementation
are acceptable---which should be true for actual ASICs where crypto cores are only a small part---or as
long as sufficient IO pins are available for an FPGA implementation. Otherwise, tuning only the FFs that are holding
the AES text is still more resilient than untuned baselines, especially in the field where other noise profiles are
coming into play as well.

\section{Conclusions and Future Work}
\label{sec:conc}

In this paper,
we have explored
joint tuning
of driver strengths and VCCs as countermeasure against PSC attacks.
Toward this end, we have proposed a simple implementation scheme, devised a CAD flow for design-time exploration of
ASICs, devised a CPA framework for thorough
and robust security analysis, and conducted a comprehensive experimental study considering both ASIC and FPGA
fabrics under various tuning scenarios.
We find that dynamic tuning is particularly effective, increasing resilience considerable for ASIC and FPGA fabrics
   along with acceptable overheads.

For future work, we plan to study joint tuning in more detail as follows.
First,
	we shall explore more efficient means for tuning, e.g., circuit-level primitives for
	ASIC implementations or an iterative strategy for IO-pin assignment for FPGA implementations.
Second,
	we shall study tuning also in the context of leakage-power attacks, given that driver strengths and
	VCCs do impact leakage-power profiles as well.
Third,
	besides using a CPA attack, we shall also utilize generic approaches for security assessment, e.g., TVLA,
	to more rigorously study possible limitations for tuning.

\balance

%\bibliographystyle{ACM-Reference-Format}
%\bibliography{main}

\end{document}